\documentclass[
twocolumn,
superscriptaddress,
 amsmath,amssymb,
 aps,
]{revtex4-2}

\usepackage{graphicx}
\usepackage{dcolumn}
\usepackage{bm}
\usepackage{hyperref}
\usepackage[mathlines]{lineno}
\usepackage{natbib}
\usepackage[utf8]{inputenc}
\usepackage{tabularx}
\usepackage{color}
\usepackage{tikz}
\usepackage[emulate=hepunits]{siunitx}
\usetikzlibrary{shapes}
\usepackage{transparent}
\graphicspath{{../figures/}}
\usepackage{tikz}
\usepackage{tikz-timing}[2009/05/15]
\usepackage{braket}
\usepackage{enumitem} 
\setlist[itemize]{itemsep = 40pt}
\usepackage{commath}

\usepackage{titlesec}
\titlespacing*{\subsection}{0pt}{1\baselineskip}{0.5\baselineskip}

\begin{document}

\newcommand{\beginsupplement}{%
        \setcounter{table}{0}
        \renewcommand{\thetable}{S\arabic{table}}%
        \setcounter{figure}{0}
		\renewcommand{\figurename}{Extended Data Fig.}
		\setcounter{equation}{0}
        \renewcommand{\theequation}{S\arabic{equation}}%
     }

\title{Observation of Feshbach resonances between a single ion and ultracold atoms}

\author{Pascal Weckesser}
\email[]{pascal.weckesser@physik.uni-freiburg.de}
\affiliation{Albert-Ludwigs-Universit\"at Freiburg, Physikalisches Institut, Hermann-Herder-Straße 3, 79104 Freiburg, Germany}
\author{Fabian Thielemann}
\affiliation{Albert-Ludwigs-Universit\"at Freiburg, Physikalisches Institut, Hermann-Herder-Straße 3, 79104 Freiburg, Germany}
\author{Dariusz Wiater}
\affiliation{Faculty of Physics, University of Warsaw, Pasteura 5, 02-093 Warsaw, Poland}
\author{Agata Wojciechowska}
\affiliation{Faculty of Physics, University of Warsaw, Pasteura 5, 02-093 Warsaw, Poland}
\author{Leon Karpa}
\affiliation{Albert-Ludwigs-Universit\"at Freiburg, Physikalisches Institut, Hermann-Herder-Straße 3, 79104 Freiburg, Germany}
\affiliation{Leibniz University Hannover, Institute of Quantum Optics, Welfengarten 1, 30167 Hannover, Germany}
\author{Krzysztof Jachymski}
\affiliation{Faculty of Physics, University of Warsaw, Pasteura 5, 02-093 Warsaw, Poland}
\author{Micha\l\, Tomza}
\affiliation{Faculty of Physics, University of Warsaw, Pasteura 5, 02-093 Warsaw, Poland}
\author{Thomas Walker}
\affiliation{Albert-Ludwigs-Universit\"at Freiburg, Physikalisches Institut, Hermann-Herder-Straße 3, 79104 Freiburg, Germany}
\author{Tobias Schaetz}
\affiliation{Albert-Ludwigs-Universit\"at Freiburg, Physikalisches Institut, Hermann-Herder-Straße 3, 79104 Freiburg, Germany}
\affiliation{EUCOR Centre for Quantum Science and Quantum Computing, Albert-Ludwigs-Universität Freiburg, Germany}

\date{\today}

\pacs{Valid PACS appear here}

\maketitle

\textbf{
Controlling physical systems and their dynamics on the level of individual quanta propels both fundamental science and quantum technologies.
Trapped atomic and molecular systems, neutral~\cite{bloch2008many} and charged~\cite{leibfried2003quantum}, are at the forefront of quantum science.
Their extraordinary level of control is evidenced by numerous applications in quantum information processing ~\cite{wineland2013nobel,saffman2010quantum} and quantum metrology~\cite{katori2011optical,micke2020coherent}.
Studying the long-range interactions between these systems when combined in a hybrid atom-ion trap~\cite{harter2014cold,tomza2019cold} has lead to landmark results~\cite{grier2009observation,Ratschbacher2012,hall2013light,ratschbacher2013decoherence,meir2016dynamics,saito2017characterization,
joger2017observation,furst2018dynamics,sikorsky2018spin,feldker2020buffer}.
Reaching the ultracold regime, however, where quantum mechanics dominates the interaction, e.g., giving access to controllable scattering resonances~\cite{idziaszek2011multichannel,tomza2015cold}, has been elusive so far.
Here we demonstrate Feshbach resonances between ions and atoms, using magnetically tunable interactions between $^{138}$Ba$^+$ ions and $^6$Li atoms.
We tune the experimental parameters to probe different interaction processes – first, enhancing three-body reactions~\cite{harter2012single,krukow2016reactive,krukow2016energy,perez2018universal} and the related losses to identify the resonances, then making two-body interactions dominant to investigate the ion's sympathetic cooling~\cite{feldker2020buffer} in the ultracold atomic bath.
Our results provide deeper insights into atom-ion interactions, giving access to complex many-body systems~\cite{cote2002mesoscopic,casteels2011polaronic,jachymski2020quantum,hirzler2020controlling} and applications in experimental quantum simulation~\cite{doerk2010atom,gerritsma2012bosonic,bissbort2013emulating}.
}

At ultracold temperatures the interaction of particles reveals its quantum mechanical nature.
For instance, considering the classical description only, a decrease of temperature would result in reduced velocities and lower reaction rates.
However, in the quantum regime , where wave-particle duality becomes dominant, a lower temperature results in an expansion of the particles' de Broglie wavelength. 
The related increase in the particles' overlap leads to an enhancement in interaction again and threshold laws predict a finite reaction rate even at zero temperature.
For small but finite energies, the quantization of angular momentum $l$ leads to the emergence of centrifugal barriers preventing the scattering into channels of higher partial waves.
This allows for a simplified view based on a small number of parameters. 
Here, the particles' wavefunction is well defined by its energy and a global phase.
In the case of elastic collisions of atoms or molecules, the kinetic energy in the centre-of-mass frame remains conserved. 
Thus, only the phase of the scattered wave function can be altered. 
It experiences a phase shift which can be related to the scattering length $a$. 
This single parameter determines cross sections and, for given densities, collision rates.
Furthermore, $a$ can be tuned and resonantly enhanced, if the free state of the colliding particles matches the energy and couples to a quasibound molecular state.
If the magnetic moments of the free and quasibound states are different, magnetic fields can be exploited to tune these into resonance using the differential Zeeman shift.
This phenomenon is known as magnetic Feshbach resonance.

During the past two decades, Feshbach resonances have become a powerful and versatile tool to control the short-range contact-interaction in homo- and heteronuclear alkali atomic mixtures~\cite{inouye1998observation,inouye2004observation}, alkali--closed-shell atomic combinations~\cite{barbe2018observation} and atom-molecule systems~\cite{yang2019observation}.
Applying Feshbach resonances to longer-range interparticle interactions has so far been limited to few highly-magnetic lanthanide atoms~\cite{aikawa2012bose,durastante2020feshbach}.
It has been proposed that the combination of atoms and ions would be a suitable candidate to introduce an isotropic, $1/r^4$ interaction potential to the quantum toolbox.
Here $r$ is the interatomic distance.
Its large characteristic range of thousands of Bohr radii~\cite{tomza2019cold}, is predicted to enable novel applications in quantum simulation~\cite{bissbort2013emulating} and quantum information processing~\cite{doerk2010atom,gerritsma2012bosonic}. 
Hybrid atom-ion systems have been widely studied and typically combine radio-frequency (rf) traps for the ion with optical dipole traps for the atoms.
However, Feshbach resonances had not been observed in such systems for two reasons.
On the one hand, the $s$-wave scattering regime for atom-ion systems is energetically at least two orders of magnitude lower than for the neutral counterpart~\cite{tomza2019cold}.
On the other hand, the ion trap's oscillating rf-fields heat the ion during elastic collisions, resulting in kinetic energies of the ion exceeding the atomic temperature by several orders of magnitude~\cite{Cetina2012,meir2016dynamics,holtkemeier2016buffer,rouse2017superstatistical}.
Combining both effects prevents the atom-ion mixture from reaching the few-partial-wave regime - a critical prerequisite for the observation of Feshbach resonances.
However, it has been conjectured~\cite{Cetina2012} that few atom-ion combinations, such as $^{138}$Ba$^+$-$^6$Li or $^{171}$Yb$^+$-$^6$Li, would be suitable to reach the few partial wave regime in a hybrid trap, as the rf-induced heating is substantially reduced for systems of large ion-to-atom mass ratio. 
Recently, collision energies allowing to enter the single-partial wave regime have been demonstrated for Yb$^+$-Li~\cite{feldker2020buffer}.

Here, we demonstrate the observation of Feshbach resonances between a single $^{138}$Ba$^+$ ion and $^6$Li atoms.
We find a total of 11 resonances and identify four of them as $s$-wave resonances.
Varying the atomic density $n$, we find three-body recombination to be the predominant loss channel for ions in the vicinity of a Feshbach resonance.
Reducing $n$ allows us to tune the atom-ion sympathetic cooling rate.
To confirm the latter, we exploit optical trapping of the ion~\cite{Lambrecht2017,schmidt2018optical,weckesser2021trapping} - paving the way for a common optical confinement of atom-ion ensembles in absence of detrimental rf-fields~\cite{schmidt2020optical}.

\begin{figure}[b!]
	\centering
  \includegraphics[width = \linewidth]{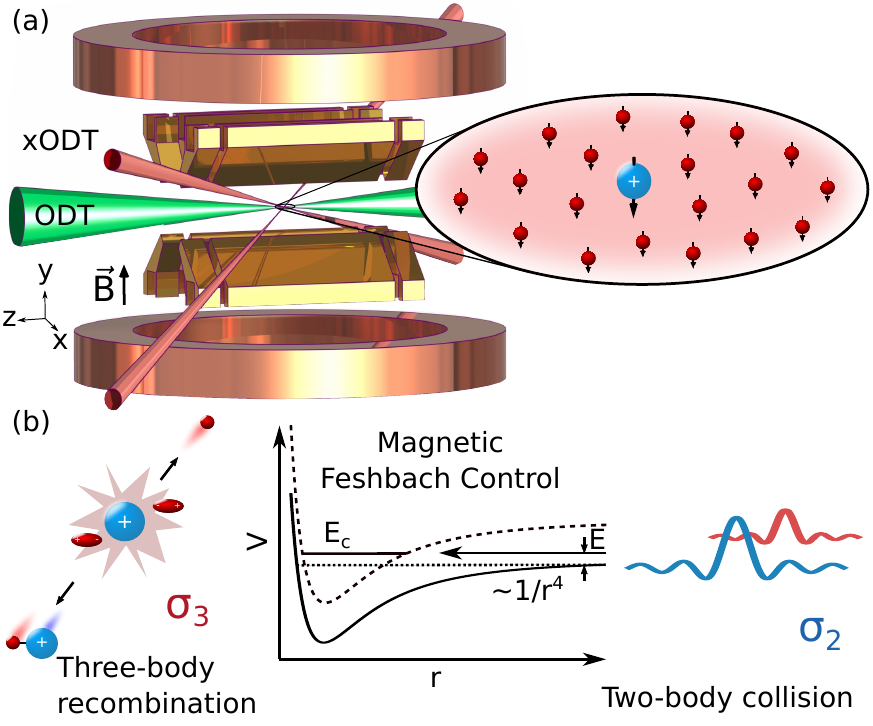}
	\caption{\textbf{Experimental setup and concepts.}
(a) We store a single $^{138}\mathrm{Ba}^+$ ion in a linear segmented rf trap, optionally in an optical dipole trap (ODT) in absence of any rf fields.
Simultaneously, we prepare an ultracold $^6\mathrm{Li}$ ensemble in a crossed ODT (xODT).
Polarizing the spins of the fermionic Li atoms suppresses mutual atom-atom interactions via the Pauli exclusion principle.
By using a set of Helmholtz-coils, we apply a homogeneous magnetic field $\vec{B}$.
(b) We control the atom-ion interactions by magnetic Feshbach resonances.
Here, two particles colliding in an open channel at energy E couple to a closed-channel molecular bound state $E_c$ of relative magnetic moment $\delta \mu$.
Tuning $B$ or choosing various $n$ and $t_{\mathrm{int}}$ allows the observation of different processes, such as elastic two-body collisions $\sigma_2$ and inelastic three-body recombination $\sigma_3$.
While the former permits sympathetic cooling of the ion, the latter can lead to the formation of weakly-bound molecular ions followed by loss.
}
	\label{fig_1}
\end{figure}

\subsection*{Experimental protocol} 

We load and confine single $^{138}\mathrm{Ba}^+$ ions deterministically in the rf trap ~\cite{weckesser2021trapping} (trap frequency $\Omega_{\mathrm{rf}} = 2\pi \times$\SI{1.433}{\mega\hertz}, secular frequencies $\omega_{\mathrm{x,y,z}}^{\mathrm{Ba}^+} = 2\pi \times \{123,122,7.6\}$\si{\kilo\hertz}).
We cool the ions close to the Doppler limit $T_{\mathrm{D}}\approx$ \SI{365}{\micro\kelvin}, compensate stray electric fields down to \SI{3}{\milli\volt/\meter} and prepare them in an incoherent spin mixture of $\ket{6S_{\mathrm{1/2}};s^{\mathrm{Ba}^+}=1/2,m_{\mathrm{s}}^{\mathrm{Ba}^+}=\pm 1/2}$ through optical pumping.
We then displace the Ba$^+$ ion from the trap centre by applying dc control fields.
Next, we create and store a cloud of $(11-33) \times 10^3$ $^6$Li atoms with temperature $T_{\mathrm{Li}}=(1-3)\, \si{\micro\kelvin}$ and density $n = (5-33)\times10^{11}\, \si{\per\centi\meter\cubed}$ in a crossed optical dipole trap (xODT) at the rf trap centre (see Methods and fig.~\ref{fig_1} (a)).
We then apply a tunable magnetic field $B$ of up to \SI{320}{\gauss} and return the ion to the trap centre where it is located inside the atomic cloud. 
We choose Li in the hyperfine state $\ket{2}=\ket{f^{\mathrm{Li}}=1/2,m_{\mathrm{f}}^{\mathrm{Li}}=-1/2}$, allowing for efficient polarization of Ba$^+$ into the $\ket{s^{\mathrm{Ba}^+} = 1/2, m_{\mathrm{s}}^{\mathrm{Ba}^+}=-1/2}$ magnetic sublevel during the first few collisions, similar to Ref.~\cite{ratschbacher2013decoherence,furst2018dynamics,sikorsky2018spin}.
For the ground state $\ket{1}=\ket{f^{\mathrm{Li}}=1/2,m_{\mathrm{f}}^{\mathrm{Li}}=+1/2}$ spin relaxation is forbidden to first order.
The ion and atoms are allowed to interact for a variable interaction duration $t_{\mathrm{int}}\in [100,300]\, \si{\milli\second}$. 
Finally, we use fluorescence detection to derive the survival probability $P_{\mathrm{ion}}$ of the Ba$^+$ in the rf trap (see Methods).
Instead of detecting the ion's state in the rf trap, we may transfer it to a single-beam optical dipole trap (ODT) for $\SI{500}{\micro\second}$, in complete absence of Li and rf-fields.
As the ODT is chosen relatively shallow, it enforces an energy cut-off for ions exceeding a dedicated kinetic energy limit.
As a result we obtain a finite optical trapping probability $p_{\mathrm{opt}}$~\cite{Schneider2012influence}, allowing to benchmark the sympathetic cooling efficiency via the atoms.
Finally, we may optionally detect the Li atom number and temperature with absorption imaging around $B_{\mathrm{Li}}\approx$\SI{293}{\gauss}.

\begin{figure*}[t!]
	\centering
  \includegraphics[width =\textwidth]{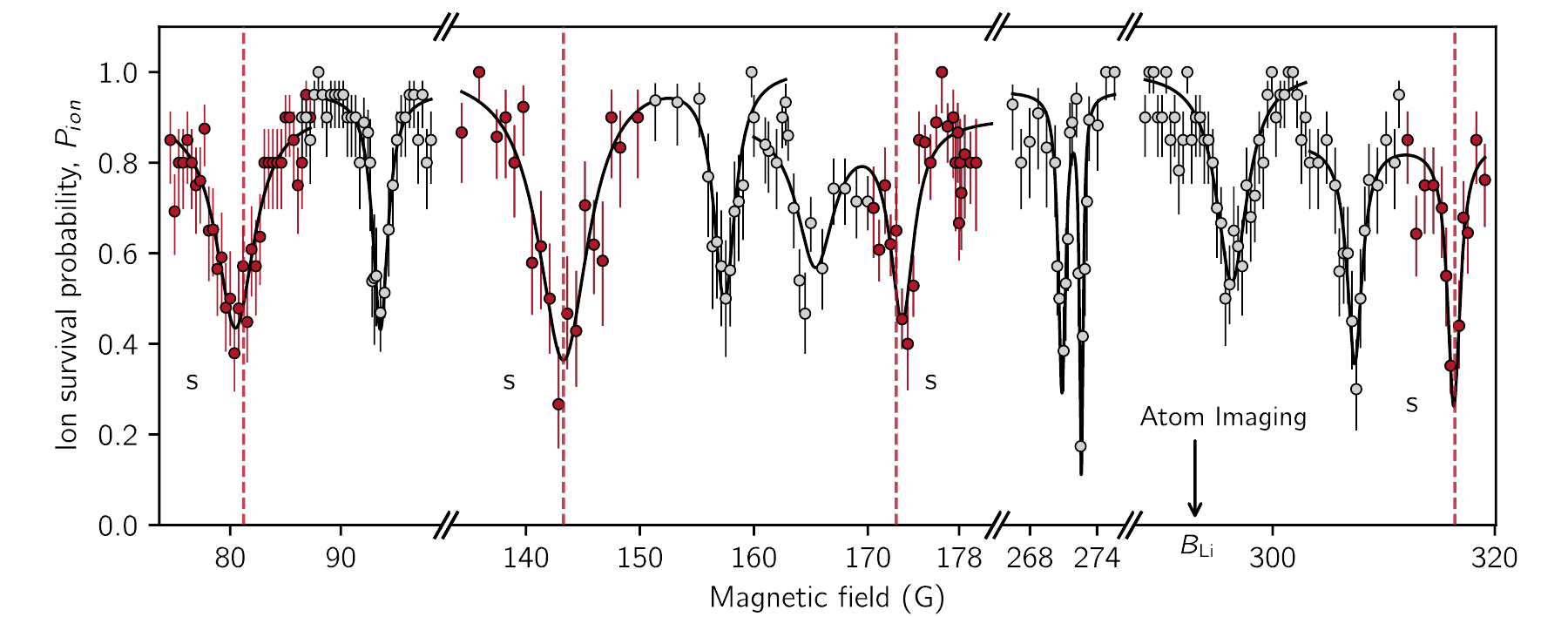}
	\caption[width = \textwidth]{\textbf{Detection of atom-ion Feshbach resonances by magnetic-field dependent ion loss spectroscopy.}
Ion survival probability for a single $^{138}\mathrm{Ba}^+$ ion embedded into $20(2)\times10^3$ spin-polarized $^6$Li atoms in dependence on $B$ and dedicated $t_{\mathrm{int}}$.
The experimental data (circles) represent an average of at least 20 experimental realizations, while the error bars denote the $1\sigma$ confidence interval.
The data was taken in random order for a given resonance.
We obtain the centre position of the resonances $B_{\mathrm{0}}^{\mathrm{expt}}$ and their FWHM  $\Delta_{\mathrm{0}}^{\mathrm{expt}}$ (see Table.~\ref{table_feshbach}) by fitting a sum of Lorentzian functions to the data points (black curve).
We assign four s-wave resonances (red data points).
Their theoretical positions $B_{\mathrm{0,s}}^{\mathrm{theo}}$ are represented by the red vertically dashed lines.
}
	\label{fig_2}
\end{figure*}

\subsection*{Locating Atom-Ion Fesbach Resonances}

As we operate with a large number of atoms ($N_{\rm atom}>10^4$) and a single ion, atom loss through atom-ion interactions is negligible.
Therefore, to perform loss spectroscopy, we search for Feshbach resonances by probing the Ba$^+$ loss, which we detect with near-unity efficiency.
Unlike in loss spectroscopy with atomic quantum gases, where the particles are irretrievably lost from the trap, the comparatively deep rf-trap can store some ionic reaction products.
The details of how these reaction products are distinguished can be found in the Methods section.

For guidance, \textit{ab initio} electronic structure calculations and multichannel quantum scattering calculations (MCQSC) are used to predict the number of resonances, their mutual distances and their approximate width $\Delta_{\mathrm{0}}^{\mathrm{theor}}$.
However, the prediction of their absolute magnetic field location $B_{\mathrm{0}}^{\mathrm{theor}}$ requires input from experimental observation.
For our system, we expect relevant contributions of partial waves for $l\leq2$ ($s$, $p$ and $d$).
MCQSC forecasts $\sim 5(1)$ Feshbach resonances, including $s$-, $p$- and $d$-wave resonances, for $B \in [70,330]\, \si{\gauss}$ with $\Delta_{\mathrm{0}}^{\mathrm{theor}} \approx \SI{1}{\gauss}$.
For this calculation we considered regular, that is, spin-projection conserving electronic interaction only ($\delta m_{\mathrm{F}} = 0$ with $m_{\mathrm{F}} = m_{\mathrm{f}}^{\mathrm{Li}} + m_{\mathrm{s}}^{\mathrm{Ba}^+}$).

In order to reveal these resonances experimentally, we performed a point-by-point search of $B_{\mathrm{0}}^{\mathrm{expt}}$ with a step-size of $\sim$\SI{400}{\milli\gauss} (see fig.~\ref{fig_2}). 
For each located resonance, we adapt $t_{\mathrm{int}}$ to obtain $P_{\mathrm{ion}}\lesssim$\SI{50}{\percent} on resonance.
Despite investigating only parts of the range $B \in [70,330]\si{\gauss}$, we already observe eleven resonances, which are summarized in table~\ref{table_feshbach}.
Note that $^6$Li prepared in spin state $\ket{2}$ with no Ba$^+$ present does not feature any resonances within the investigated $B$-field ranges.

\begin{table}[b!]
  \begin{center}
    \begin{tabular*}{0.9\linewidth}{c @{\extracolsep{\fill}}  c | c |  c  } 
    \hline
    \hline
      $B^{\mathrm{expt}}_{\mathrm{0}}$ (\si{\gauss})  & $\Delta^{\mathrm{expt}}_{\mathrm{0}}$ (\si{\gauss}) & $B^{\mathrm{theo}}_{\mathrm{0,s}}$(\si{\gauss}) & $t_{\mathrm{int}}$ (\si{\milli\second})  \\
      \hline
       $80.47(21)$ & $4.2(1.1)$ & 81.2 &   300 \\
       $93.59(9)$ & $1.6(4)$ &  & 100 \\
       $143.29(31)$ & $5.3(1.3)$  & 143.3 & 300 \\
       $157.42(19)$ & $2.2(7)$  &   & 300 \\
       $165.4(4)$ & $4.0(1.1)$  &   & 300 \\
	   $173.0(2)$ & $2.5(5)$ & 172.5  &  300 \\
       $270.86(5)$ & $0.76(24)$ & &  300 \\
       $272.59(3)$ & $0.4(1)$ & &  300 \\  
       $296.31(19)$ & $3.4(7)$ & &  300 \\       
       $307.38(14)$ & $1.6(5)$ & & 200 \\
       $316.31(11)$  & $1.3(5)$ & 316.4 & 200 \\      
      \hline
      \hline
    \end{tabular*}
  \end{center}
 \caption{\textbf{List of observed Feshbach resonances for the entrance channel $\mathbf{^{138} \text{Ba}^+ \ket{s^{\mathrm{Ba}^+} = 1/2, m_{\text{s}}^{\mathrm{Ba}^+} = - 1/2}}$ $\mathbf{+}$ $\mathbf{^6 \text{Li}\ket{ f^{\mathrm{Li}} = 1/2, m_{\text{f}}^{\mathrm{Li}} = -1/2}}$.}
The presented errors equal the $1\sigma$ confidence interval of the fit.
We can further assign an overall systematic error of $60\,\mathrm{mG}$ on $\delta B_{\mathrm{0}}^{\mathrm{expt}}$, including daily drifts and calibration uncertainties.
We compare our measurements with an ABM and can assign four s-wave resonances $B^{\mathrm{theo}}_{\mathrm{0,s}}$ for $a_{\mathrm{S}} = 0.236\, R_{\mathrm{4}}$ and $a_{\mathrm{T}} = -0.053\, R_{\mathrm{4}}$.
}
 \label{table_feshbach}
\end{table}

\subsection*{Theoretical resonance assignment} 

The observed number of resonances (11) exceeds the initial prediction by MCQSC ($\sim 5(1)$).
This indicates that additional coupling terms besides regular electronic interaction have to be considered, such as the second-order spin-orbit coupling (SOC).
Since SOC can substantially mix the internal spins ($m_{\mathrm{F}}$) with the rotational motion ($l$, $m_l$), the number of resonances would be increased~\cite{ticknor2004multiplet}.
Note that $m_{\mathrm{F}}$-changing couplings can also split and shift higher-partial wave resonances~\cite{ticknor2004multiplet,cui2017observation}, while $s$-wave resonances remain unaffected to first order.
Here we use the term $l$-wave resonances if the entrance collision channel is coupled to an $l$-wave molecular level.

As an initial step towards a complete partial-wave assignment, we first attempt to identify the $s$-wave resonances with $\abs{\delta m_{\mathrm{F}}} = \{0, 1, 2\}$, the positions of which depend only on the singlet and triplet scattering lengths $a_{\mathrm{S}}$ and $a_{\mathrm{T}}$. 
The first 9 discovered resonances were used to obtain best-fit values for $a_{\mathrm{S}}$ and $a_{\mathrm{T}}$ from the asymptotic bound-state model (ABM; see Methods).
From this, three $s$-wave resonances (all $\delta m_{\mathrm{F}} =1$) were identified and an additional one with $\delta m_{\mathrm{F}} =1$ was predicted around $B_{\mathrm{0,s}}^{\mathrm{theor}}=\SI{172}{\gauss}$.
Subsequently, we observed a resonance at $B_{\mathrm{0}}^{\mathrm{expt}} = \SI{173.0(2)}{\gauss}$, being a strong evidence for correct $s$-wave assignment.
We find $a_{\mathrm{S}} = 0.236\, R_{4}$ and $a_{\mathrm{T}} = -0.053\, R_{4}$ ($R_4=\SI{69}{\nano\meter}$) using all 4 $s$-wave resonances (see Methods).
Note that one can derive the scattering length $a$ out of $a_{\mathrm{S}}$ and $a_{\mathrm{T}}$. 
The presented accuracy of the assignment is comparable to those typically achieved for Feshbach resonances of neutral atoms.
To provide an assignment for $l\geq 1$, a refined model is currently under investigation.

\subsection*{Atomic density-dependent ion loss}

\begin{figure}[t!]
	\centering
  \includegraphics[width =  \linewidth]{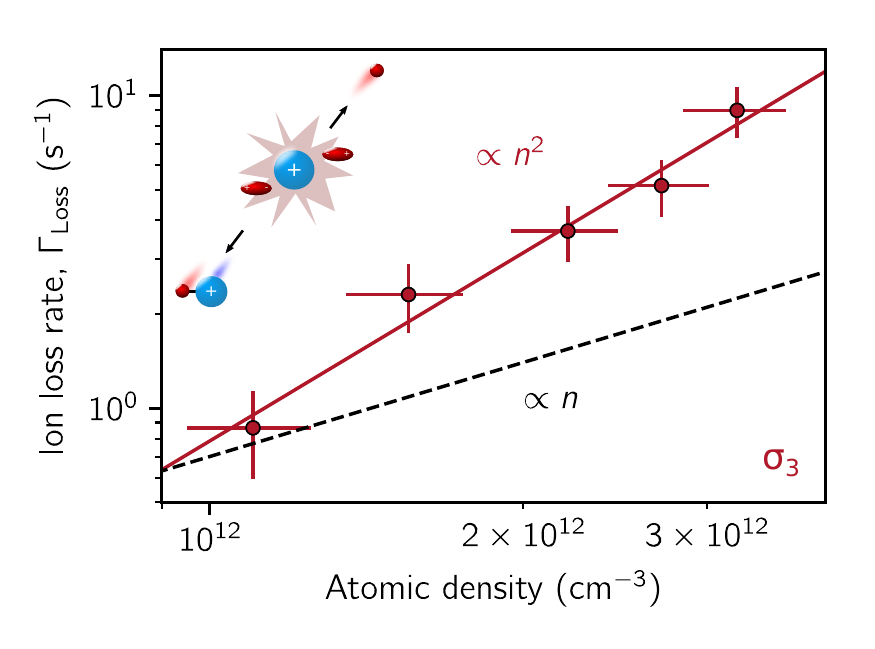}
	\caption{\textbf{Ion loss rate in dependence on the atomic density at the Feshbach resonance at 296.31 G.}
The vertical errors represent the $1\sigma$ confidence interval of the fit.
The horizontal errors represent the systematic uncertainty of the atomic density, due to fluctuations in atom-ion overlap and atom number.
The red solid line shows the best fit for $\Gamma_{\mathrm{Loss}}(n) = k_{\mathrm{2}}n +  k_{\mathrm{3}}n^2 \approx k_{\mathrm{3}}n^2$.
To emphasize the deviation of a linear dependency on $n$, which is the expected scaling for two-body loss channels, we additionally present a linear scaling (dashed black line).
}
	\label{fig_3}
\end{figure}

Independent of the current assignment, we further examine the nature of the Ba$^+$ loss processes and their dynamics.
To study the number of atoms involved in the loss process of the ion and the related timescales, we probe $P_{\mathrm{ion}}$ in dependence on $t_{\mathrm{int}}$ for five different $n$.
We selected the Feshbach resonance at \SI{296.31}{\gauss}, as it lies closest to $B_{\mathrm{Li}}$.
Here we derive $n$ with highest accuracy, while we avoid larger changes of the $B$-field, potentially sweeping over atom-ion resonances not yet identified.
For each time evolution, we fit an exponential decay $e^{-t_{\mathrm{int}}\Gamma_{\mathrm{Loss}}}$ (see Extended Data).

The results for $\Gamma_{\mathrm{Loss}}(n)$ are illustrated in fig.~\ref{fig_3}.
We model the data by $\Gamma_{\mathrm{Loss}}(n) = k_{\mathrm{2}}n +  k_{\mathrm{3}}n^2 $, with $k_{\mathrm{2}}$ and $k_{\mathrm{3}}$ representing the two- and three-body loss rate coefficients.
We derive $k_{\mathrm{3}}=7.8(1.3)_{\mathrm{stat}}(1.1)_{\mathrm{sys}} \times 10^{-25}\,\si{\centi\meter^6\per\second}$ while $k_{\mathrm{2}}$ remains consistent with zero.
Alternatively, we fit the exponent $c$ of $\Gamma_{\mathrm{Loss}}(n)\propto n^c$, resulting in $c=2.02(29)$.
We conclude that the ion losses are dominated by three-body recombination (TBR)~\cite{harter2012single,krukow2016reactive,krukow2016energy,perez2018universal}.
The increase of TBR in the vicinity of a Feshbach resonance is directly related to the enhancement of two-body interaction.
Note that TBR occurs despite operating with a spin-polarized fermionic atomic ensemble obeying the Pauli exclusion principle for Li-Li interactions.

The process of TBR and the related width of a resonance is strongly temperature-dependent~\cite{maier2015emergence}.
Resolving Feshbach resonances down to a FWHM of \SI{0.4(1)}{\gauss}, we can estimate an upper bound for the collision energy $E_{\mathrm{col}}$ in the two-body centre-of-mass frame.
Here we assume that $\Delta_{\mathrm{0}}^{\mathrm{expt}}$ solely results from temperature broadening and we take the maximal differential magnetic moment for a two-valence-electron system ($\delta \mu = 4\mu_{\mathrm{B}}$).
This coarse estimate based on the Breit-Wigner formula indicates that such narrow resonances only become detectable for $E_{\mathrm{col}}\lesssim \SI{50}{\micro\kelvin} \times k_{\mathrm{B}}$, regardless of their partial wave character.
Note that $E_{\mathrm{col}}$ is below the centrifugal barrier of $d$-wave scattering ($E_d = \SI{77.4}{\micro\kelvin}\times k_{\mathrm{B}}$).

\subsection*{Magnetic tunability of cross sections} 

To study the predicted versatility of atom-ion Feshbach resonances, we investigate the prospects of controlling our experimental parameters to mitigate TBR losses while enhancing the sympathetic cooling of the Ba$^+$ ion by elastic two-body collisions with the Li atoms.
For a direct comparison of $\sigma_2$ and $\sigma_3$, we again operate close to $B_{\mathrm{0}}^{\mathrm{expt}}=\SI{296.31}{\gauss}$. 
We decrease $n$ and $t_{\mathrm{int}}$ to $5.5(2.0)\times 10^{11}\,\si{\per\cubic\centi\meter}$ and $\SI{150}{\milli\second}$, resulting in the ion survival probability $P_{\mathrm{ion}}$ of $95.7^{+2.7}_{-6.5}\si{\percent}$ for identical rf-confinement in the hybrid trap. 
Under these conditions, we probe the dependence of sympathetic cooling efficiency on $B$.
Following the interaction with the atoms, we transfer the Ba$^+$ ion into the ODT of trap depth $U_{\mathrm{0}}^{\mathrm{ODT}}=\SI{620(113)}{\micro\kelvin}\times k_{\mathrm{B}}$.
Note that after the interaction the actual energy distribution of the ion deviates from an ideal Boltzmann distribution, e.g., as the ion has not reached its kinetic steady state for the chosen $t_{\mathrm{int}}$ and as rf-heating effects typically result in non-thermal Tsallis distributions~\cite{rouse2017superstatistical}.
As a consequence, we cannot derive a temperature representing the ion's kinetic energy $E_{\mathrm{Ba}^+}$ as presented in ref.~\cite{Schneider2012influence}.
Nevertheless, the optical trapping probability $p_{\mathrm{opt}}$ is directly related to $E_{\mathrm{Ba}^+}$, that is, for lower $E_{\mathrm{Ba}^+}$ we expect higher $p_{\mathrm{opt}}$ and vice versa.

\begin{figure}[t!]
	\centering
  \includegraphics[width =  \linewidth]{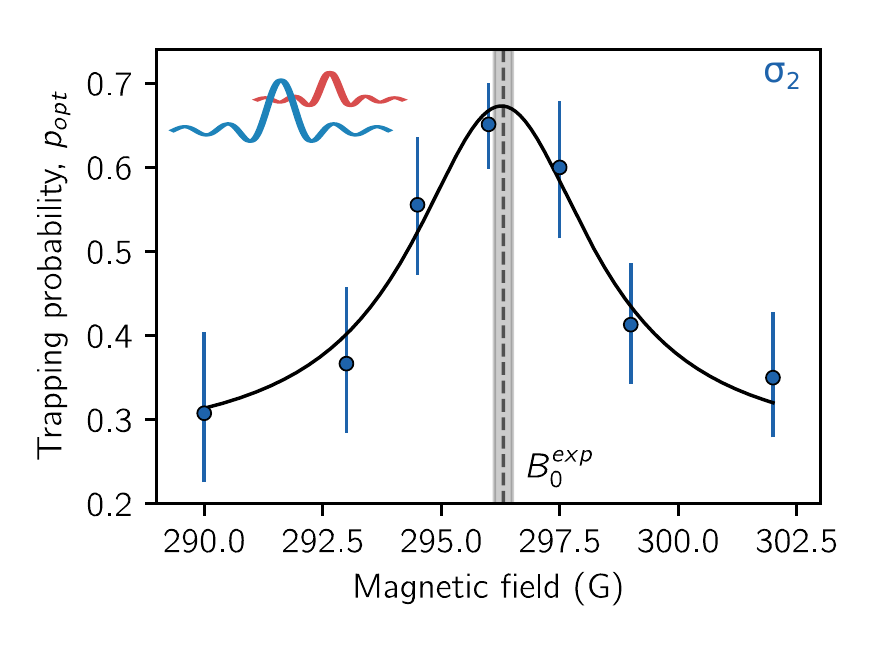}
	\caption{\textbf{Enhanced sympathetic cooling in vicinity of the Feshbach resonance at 296.31 G.}
After interacting for $t_{\mathrm{int}} = \SI{150}{\milli\second}$ with the Li atoms, we transfer the Ba$^+$ ion into the shallow ODT with fixed trap depth $U_{\mathrm{0}}^{\mathrm{ODT}} = 620(113)$\si{\micro\kelvin}$\times k_{\mathrm{B}}$ and record the optical trapping probability $p_{\mathrm{opt}}$ in dependence on $B$.
The data points are an average of up to 86 independent realizations and were recorded in random order.
The error bars denote the $1\sigma$ confidence interval.
In the vicinity of the Feshbach resonance at $B_{\mathrm{0}}^{\mathrm{expt}} = \SI{296.31(19)}{\gauss}$ (vertical dashed line with respective error bars denoted as shaded area) we observe a maximum for $p_{\mathrm{opt}}$, being strong evidence for enhanced sympathetic cooling due to an increase of the elastic cross section $\sigma_2$.
Fitting the data with a phenomenological Lorentzian function (black solid curve), results in a centre position \SI{296.3(2)}{\gauss} and FWHM of $4.6(1.1)\,\si{\gauss}$, being consistent to $B_{\mathrm{0}}^{\mathrm{expt}}$.
}
	\label{fig_4}
\end{figure}

Running the experimental protocol in absence of atom-ion interaction, we observe $p_{\mathrm{opt}}<\SI{5}{\percent}$. 
Operating with atom-ion interaction detuned by $\abs{B-B_{\mathrm{0}}^{\mathrm{expt}}}>\Delta_{\mathrm{0}}^{\mathrm{expt}}$ from resonance we record $p_{\mathrm{opt}}\sim\SI{30}{\percent}$ (see fig.~\ref{fig_4}).
Tuning to $B_{\mathrm{0}}^{\mathrm{expt}}$, we observe a further increase towards $p_{\mathrm{opt}}\approx\SI{65}{\percent}$.
We see this as evidence for enhanced sympathetic cooling related to an increase of the elastic two-body cross section $\sigma_2(B)$, a key attribute of Feshbach resonances.

\subsection*{Conclusions and future perspectives}

We have observed magnetically tunable Feshbach resonances between a single ion and an ultracold atomic gas.
Depending on the atomic density, we found the tunable atom-ion interactions to be either dominated by two- or three-body collisions.

Moving beyond this first demonstration, we will extend the control over the interactions of choice by coherently controlling the electronic degrees of freedom of $^{138}$Ba$^+$ including deterministic state-preparation and investigating $^6$Li in its hyperfine ground state.
Further optimizing the sympathetic cooling might open the door to an all-optical approach~\cite{schmidt2020optical}, where the ion(s) and atoms are confined in an optical trap simultaneously, in absence of disturbing rf-fields and their detrimental heating mechanisms.
This is predicted to permit studying chemistry in the $s$-wave regime, where molecular ions can be formed coherently by magneto- or rf-association~\cite{kohler2006production}, as well as stored and analyzed after exothermic reactions, e.g. via deep box-shaped rf potentials~\cite{wester2009radiofrequency}. 
Further experimental data on certain resonances will allow for improvements to the accuracy and detail of predictions from MCQSC. 
This might provide deeper insight into the molecular structure at short-range, where additional interactions, such as SOC, shift and split resonances~\cite{ticknor2004multiplet}.
Furthermore, exploiting Feshbach resonances in a generic atom-ion ensemble might find application in many-body systems, such as polaron~\cite{casteels2011polaronic,jachymski2020quantum} and impurity physics~\cite{hirzler2020controlling} as well as experimental quantum simulations~\cite{doerk2010atom,gerritsma2012bosonic,bissbort2013emulating}.

\bibliography{Feshbach}


\section{Methods}

\textbf{Preparation of spin-polarized $^6$Li ensemble.}
To prepare our $^6$Li ensemble, we first load a magneto-optical trap, and then transfer the atoms into a crossed optical dipole trap (xODT).
During the transfer into the xODT, we optically pump the Li atoms into the lowest two hyperfine states $\ket{1}=\ket{f^{\mathrm{Li}}=1/2,m_{\mathrm{f}}^{\mathrm{Li}}=+1/2}$ and $\ket{2}=\ket{f^{\mathrm{Li}}=1/2,m_{\mathrm{f}}^{\mathrm{Li}}=-1/2}$, producing an incoherent $50/50$ mixture. 
We then switch to higher magnetic fields $B\approx\SI{345.9}{\gauss}$ to increase the elastic cross section between the $^6$Li atoms~\cite{jochim2002magnetic}.
Gradually lowering the power of the xODT over \SI{900}{\milli\second} as in Ref.~\cite{Luo2006}, we evaporatively cool the atomic cloud and obtain approximately $40\times 10^3$ $^6$Li atoms with final temperatures on the order of $(1-3)\,\si{\micro\kelvin}$.
For the lowest temperatures we obtain typical trapping frequencies $\omega_{x,y,z}^{\mathrm{Li}} =2\pi\times \{1.77(10),1.94(10),0.19(1)\}\,\mathrm{kHz}$.
Finally, we prepare the Li ensemble in spin state $\ket{2}$ through spin-selective absorption imaging of state $\ket{1}$ for \SI{100}{\micro\second} at $B\approx\SI{345.9}{\gauss}$.
We detect the Li atoms in spins state $\ket{2}$ by applying high-field absorption imaging at $B \sim \SI{293}{\gauss}$.
We can further determine the temperature of the atomic ensemble $T_{\mathrm{Li}}$ by time-of-flight thermometry~\cite{ketterle2008making}. 

\textbf{The crossed dipole trap (xODT).}
The xODT is created by a commercial solid-state laser at \SI{1064}{\nano\meter} with a maximum output power of $\SI{50}{\watt}$.
The trap consists of two linearly-polarized laser beams with orthogonal polarization crossing at an angle of $\sim \ang{14}$.
Here both $k$-vectors intersect the $\hat{x}$-$\hat{z}$-plane and the ion trap's principal axis $\hat{z}$ at an angle of $\sim \ang{7} $ and $\sim \ang{31} $, respectively.
The foci of the two beams are overlapped and the beam waists are \SI{30(3)}{\micro\meter}.
We superimpose the centre of the xODT with the ion by maximizing the ac Stark shift caused by the xODT on the ion.
We validate the atom-ion overlap with an independent reference measurement.
Similar to Ref.~\cite{joger2017observation}, we probe the atomic density profile by recording the ion loss rate for a Ba$^+$ ion being prepared in the $5$D$_{\mathrm{3/2}}$ manifold.
We find consistency between the two methods. 

\textbf{Optical trapping of $^{138}$Ba$^+$ ions in the ODT.}
After the atom-ion interaction, we can transfer the Ba$^+$ ion into the single beam optical dipole trap (ODT).
The ODT is operated at \SI{532}{\nano\meter} and propagates along the $\hat{z}$-axis of the rf trap ($\hat{z}$).
The beam waists equal $w_x=\SI{4.5(2)}{\micro\meter}$ and $w_y=\SI{4.0(2)}{\micro\meter}$ at the location of the ion.
Note that the ODT acts attractively for Ba$^+$ prepared in the $6\mathrm{S}_{1/2}$ manifold~\cite{weckesser2021trapping}, while it is blue-detuned, and thus repulsive, for the Li atoms.
Independent calibration measurements show that the presence of the atoms does not influence the ion's optical trapping probability $p_{\mathrm{opt}}$, as the repulsive ODT separates the ensemble from the ion on timescales of $<\SI{100}{\micro\second}$.
Transferring the ion into the ODT and turning off the rf fields follows the experimental protocol as in Ref.~\cite{weckesser2021trapping}.
At the end, we transfer the ion back into the linear rf trap and detect it via fluorescence imaging while Doppler cooling.
If we detect an ion in the rf trap, we consider the optical trapping attempt successful.
If we observe no ion, the ion was irretrievably lost from the ODT and the attempt was unsuccessful.
Repeating the protocol, we obtain the optical trapping probability $p_{\mathrm{opt}}$. 

\textbf{Calculating $\mathbf{U_{\mathrm{0}}^{ODT}}$.}
Unlike for neutral atoms where the potential trap depth predominantly depends on the ac Stark shift~\cite{grimm2000optical}, the Ba$^+$ ion is subject to the Coulomb interaction and therefore very sensitive to electric fields~\cite{karpa2019trapping}.
The electro-optical potential is therefore given by three contributions: (1) the ac Stark shift, (2) residual stray electric fields and (3) dc defocusing curvatures.
During an optical trapping attempt, we confine the ion along the ODT's $\vec{k}$-vector ($\vec{k}\parallel \hat{z}$) using dc contributions.
Note that an axial dc confinement inevitably results in defocussing in the radial $\hat{x}$-$\hat{y}$-plane~\cite{maxwell1873treatise}.
Independent calibration measurements reveal radial stray electric fields and radial defocussing dc curvatures of $E_{\mathrm{S}}\leq \SI{3}{\milli\volt/\meter}$ and $m \omega_{dc}^2$ with $\omega_{\mathrm{dc}}^2 =  -  (2\pi\times 6.5(1))^2\,\si{\kilo\hertz}^2$, respectively.
Knowing the beam waist and the electrical contributions, we derive the electro-optical trap depth $U_{\mathrm{0}}^{\mathrm{ODT}}$ as presented in Ref.~\cite{weckesser2021trapping}. 

\textbf{Probing the ion survival probability $\mathbf{P_{\mathrm{\textbf{ion}}}}$.}
During the atom-ion interaction, the Ba$^+$ ion might undergo various processes in the hybrid trap.
The ion can collide elastically, meaning it remains in the $6$S$_{\mathrm{1/2}}$ electronic ground state while exchanging energy with the Li atoms, acting as cooling agent.
In contrast, the ion might undergo an inelastic spin-changing collision or three-body recombination (TBR).
The latter predominantly results in the formation of weakly-bound molecular ions~\cite{perez2015communication,perez2018universal} followed by loss.
The losses can be partially explained by a combination of collisional quenching and light-assisted dissociation.
For the latter the ion can gain significant kinetic energy, change its electronic state or undergo a non-radiative charge-exchange resulting in neutral Ba and in charged Li$^+$.
Except for the case of charge-exchange, the Ba$^+$ ion remains within the deep trapping volume of the Paul trap and its presence can be detected by fluorescence imaging.
Similar observations have recently been reported for Ba$^+$-Rb~\cite{mohammadi2021life}.
However, the exact mechanism remains to be investigated for Ba$^+$-Li.

\begin{figure}[]
	\centering
  \includegraphics[width =  \linewidth]{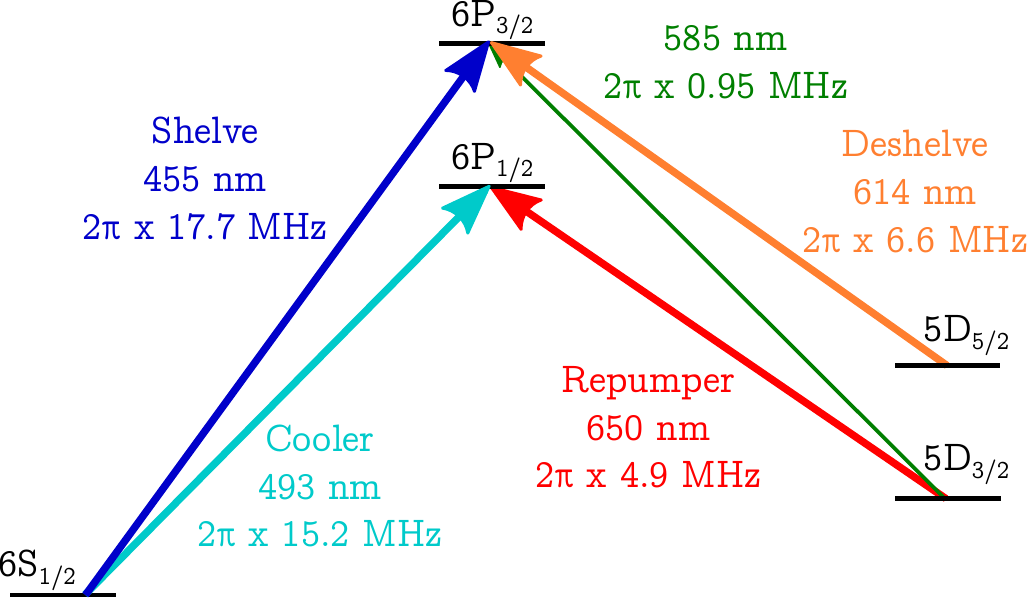}
	\caption{\textbf{Electronic level scheme of $\mathbf{^{138}}$Ba$^{+}$ ($I=0$).}
We label the relevant electronic dipole transitions with their respective wavelength $\lambda$ and natural linewidth $\Gamma$.	
We Doppler cool the ion driving the $6\mathrm{S}_{\mathrm{1/2}} \leftrightarrow 6\mathrm{P}_{\mathrm{1/2}}$ and $5\mathrm{D}_{\mathrm{3/2}} \leftrightarrow 6\mathrm{P}_{\mathrm{1/2}}$ transition.
Inelastic losses, such as three-body recombination followed by light-assisted dissociation, can partially result in the ion's population of the $5$D$_{5/2}$ manifold~\cite{mohammadi2021life}.
We detect these events through optical pumping with $\SI{614}{\nano\meter}$ laser light, followed by fluorescence detection while Doppler cooling.
}
	\label{suppl_fig_1}
\end{figure}

In order to reliably discriminate between elastic and inelastic collisions, we apply the following protocol.
In a first step, we apply Doppler cooling using laser light at \SI{493}{\nano\meter} and \SI{650}{\nano\meter} (see energy levels and related transitions in fig.~\ref{suppl_fig_1} in Extended Data).
Here, we operate the \SI{493}{\nano\meter} cooling laser close to resonance, with a detuning of $\delta = -2/3 \Gamma$ with $\Gamma = 2 \pi \times \SI{15.2}{\mega\hertz}$ being the natural linewidth of the excited $6P_{\mathrm{1/2}}$ state.
For this detuning, we can detect cold Ba$^+$ ions populating the $6$S$_{\mathrm{1/2}}$ manifold. 
If detected, we consider this a survival-event where the ion interacted purely elastically with the Li atoms.
In order to detect Ba$^+$ of increased $E_{\mathrm{kin}}$, e.g. due to photodissociation ($>\SI{100}{\kelvin}\times k_{\mathrm{B}}$), we apply an additional far red-detuned cooling laser with $\delta \approx -14 \Gamma$ for \SI{1}{\second}.
This laser is able to cool the ions of increased $E_{\mathrm{kin}}$ into the sub-Kelvin regime, where they become detectable by the nearer-detuned fluorescence imaging beam.
Taking a second fluorescence image, we detect those ions.
As final measurement, we examine whether the ion populates the metastable $5$D$_{\mathrm{5/2}}$ manifold.
Here, we apply laser light at \SI{614}{\nano\meter}, optically pumping the ion into the electronic ground state via the $6$P$_{3/2}$ state (see fig.~\ref{suppl_fig_1} in Extended Data).
Taking a third fluorescence image, we can detect these ions.
In the case that we do not detect any ion, we assume the Ba$^+$ to have undergone charge-exchange.
For the ion loss spectroscopy in fig.~\ref{fig_2}, the ion survival probability $P_{\mathrm{ion}}$ is defined by the number of survival-events divided by the total number of events (elastic and inelastic).
Inelastic events are the sum of ions of increased $E_{\mathrm{kin}}$, ions populating the $5$D$_{\mathrm{5/2}}$ manifold and non-detectable ions. 

\textbf{Magnetic field calibration.}
Magnetic field calibration is achieved by driving a resonant magnetic-field dependent nuclear spin transition in $^6$Li between the states $\ket{1}\leftrightarrow \ket{2}$.
We determine and benchmark the magnetic fields using the Breit-Rabi formula \cite{breit1931measurement}.
The calibration was carried out in the vicinity of all recorded Feshbach resonances from table~\ref{table_feshbach}.
Here we sampled the entire magnetic field range of $[80,320]\si{\gauss}$ with equidistant data points and interpolated intermittent values by a global linear fit function.
We derive a systematic magnetic field uncertainty of \SI{60}{\milli\gauss}, including day-to-day drifts and minor deviations due to the interpolation.
As an independent validation, we probed the $\ket{1} - \ket{1}$ and $\ket{2} - \ket{2}$ atomic p-wave Feshbach resonance in $^6$Li around \SI{159}{\gauss} and \SI{215}{\gauss} respectively, finding agreement with previous publications \cite{zhang2004p}.

\textbf{List of scanned magnetic field ranges.}
During our investigation, we scanned the following magnetic field ranges: [5,100]\si{\gauss}, [118,190]\si{\gauss}, [265,280]\si{\gauss} and [285,320]\si{\gauss}.
Fig.~\ref{fig_2} illustrates the magnetic field windows of the ranges within which we identified Feshbach resonances.
For the remaining parts, we did not record any resonance with significant signal-to-noise ratio.
Here we typically scanned magnetic subsets of up to \SI{10}{\gauss} on one day.

\textbf{Electronic structure calculations.}
Potential energy curves for the singlet $X^1\Sigma^+$ and triplet $a^3\Sigma^+$ molecular electronic states, resulting from the interaction between a ground-state Ba$^+$ ion and a Li atom, are calculated using \textit{ab initio} electronic structure approaches, such as the coupled cluster method restricted to single, double, and noniterative triple excitations and the multireference configuration interaction method restricted to single and double excitations with aug-cc-pwCV5Z basis sets and ECP46MDF small-core pseudopotential for Ba$^+$ to include scalar relativistic effects.. Computational scheme following Ref.~\cite{tomza2015cold} is employed.
 
The crossing between the $a^3\Sigma^+$ and $b^3\Pi$ molecular electronic states below the collision atomic threshold is predicted (see Extended Data). It provides a mechanism for large spin-orbit mixing and large second-order spin-orbit interaction responsible for strong spin-nonconserving scattering~\cite{mies1996estimating,tscherbul2016spin}.    

\textbf{Multichannel quantum scattering calculations.}
Two-body collisions are studied using numerically exact multichannel quantum scattering calculations in free space~\cite{tomza2019cold}. The Hamiltonian used for the nuclear motion includes the singlet and triplet molecular electronic states, the molecular rotation, the hyperfine and Zeeman interactions, and the interatomic spin-spin interaction resulting from the magnetic dipolar and second-order spin-orbit couplings. The total scattering wave function is constructed in a complete basis set containing electronic spins, nuclear spins, and rotational angular momenta. Experimental values of atomic parameters are assumed. The scattering lengths of the calculated singlet and triplet potentials are adjusted by applying uniform scaling factors $\lambda_i$ to the interaction potentials: $V_i(r)\to \lambda_iV_i(r)$. The scattering lengths are expressed in units of the characteristic length scale for the atom-ion interaction $R_4=\sqrt{\mu C_4/\hbar}$. The coupled-channel equations are solved using QDYN~\cite{tomza2015cold} based on $S$-matrix formalism.
Our theoretical approach neglects the impact of the micromotion, trap confinement, and Lorentz force on atom-ion scattering. Therefore, the assigned singlet and triplet scattering lengths should be considered as effective ones including shifts by neglected effects.

\textbf{Asymptotic-bound-state model.}
The spectrum of bound molecular levels and positions of Feshbach resonances are obtained using the asymptotic-bound-state model (ABM)~\cite{wille2008exploring,tiecke2010asymptotic}. The same Hamiltonian and wave-function representation as in the multichannel quantum scattering calculations are employed. The ABM model is used for straightforward and fast assignment of resonance positions and nature of underlying molecular levels.

\textbf{Assignment of resonances.}
The positions of Feshbach resonances, $B_i^\text{theo}$, are calculated for a set of $10^4$ combinations of the singlet, $a_S$, and triplet, $a_T$, scattering lengths within the ABM model. A grid equidistant in the scattering phase shifts between $-\pi/2$ and $\pi/2$ is used. Resonances between the collision channel $m_F=-1$ and all coupled channels are located.

Next the predicted theoretical positions of Feshbach resonances are compared with measured positions $B_i^\text{expt}$. The singlet and triplet scattering lengths corresponding to experimental data are derived by minimizing the $\chi^2$ function
\begin{equation}
\chi^2(a_S,a_T)=\sum_{i=1}^{N_\text{expt}} \left(B_i^\text{expt}-B_i^\text{theo}(a_S,a_T)\right)^2\,,
\end{equation}
which quantifies how well our theoretical model reproduces the measured positions for $N_\text{expt}$ datapoints. The related root-mean-square deviation $RMSD=\sqrt{\chi^2/N_\text{expt}}$ can also be used to measure agreement between theory and experiment. 

For the initial assignment, we used resonances resulting from coupling to $s$-wave molecular levels because such resonances are not affected by the spin-orbit coupling in the first order of perturbation theory. Initially, 9 loss features at 80.47, 93.59, 143.29, 157.42, 270.86, 272.59, 296.31, 307.41, 316.31 were measured, which were more numerous than number of possible $s$-wave resonances in studied magnetic field range. Therefore, we calculated $\chi^2$ for all possible subsets of measured features and all combinations of scattering lengths. 2-element subsets were too small to conclude any assignment, while 4 and more element subsets could not be described by the used theoretical model. Among a few possible 3-element sets, only one would require the existence of an additional $s$-wave resonance around 172$\,$G, which next was measured around the predicted position. Finally, we found the best agreement for 4 loss features at the 80.47, 143.29, 173.0, 316.31 G reproduced as $s$-wave resonances by the ABM model with $\chi^2=0.78\,$G$^2$ and $RMSD=0.44\,$G. The found optimal scattering lengths were $a_S=0.236\, R_4$ and $a_T=-0.053\, R_4$. The second relatively small $\chi^2$ for 4 $s$-wave resonances was 3.5 times larger than the optimal one. Other possible assignments for larger number of $s$-wave resonances would result in significantly larger $\chi^2$.

The existence of higher partial-wave resonances was neglected in this assignment but similar agreement for higher-wave resonances was obtained with the optimal set of the scattering lengths, while other assignments were not able to reproduce all measured features. A detailed study of higher partial-wave resonances will be presented in the future. 
 

\bigskip
\textbf{Acknowledgements.}
This project has received funding from the European Research Council (ERC) under the European Union’s Horizon 2020 research and innovation program (Grant No. 648330) and was supported by the Georg H. Endress foundation.
P.W., F.T. and T.S. acknowledge support from the DFG within the GRK 2079/1 program.
P.W. gratefully acknowledges financial support from the Studienstiftung des deutschen Volkes.
L. K. is grateful for financial support from Marie Curie Actions.
D.W., A.W., and M.T. acknowledge the financial support from the National Science Centre Poland (Grants No. 2016/23/B/ST4/03231 and
2020/36/T/ST2/00591) and Foundation for Polish Science within the First Team program co-financed by the European Union under the European Regional Development Fund.
K.J. acknowledges support from the Polish National Agency for Academic Exchange (NAWA) via the Polish Returns 2019 programme.
The computational part was partially supported by the PL-Grid Infrastructure.
We thank Olivier Dulieu for fruitful discussions.
We thank Markus Debatin for contributing building the experimental apparatus.

\bigskip
\textbf{Data availability.}
The experimental data that support the findings of this study are available from the corresponding author upon reasonable request.

\bigskip
\textbf{Author contributions.}
T.S. conceived the experiments.
P.W. and F.T. contributed equally to the construction of the setup, carrying out of the experiments, discussion of the results and analysis of the data and were supported by L.K. and T.W.
D.W., A.W., K.J., and M.T. performed theoretical calculations and analysis supervised by M.T.
T.S. supervised the work.
P.W. and T.S. wrote the manuscript with contributions from T.W., K.J. and M.T.
All authors worked on the interpretation of the data and contributed to the final manuscript.

\bigskip
\textbf{Competing interests.}
The authors declare no competing interests.

\beginsupplement

\newpage

\,

\newpage

\begin{figure}[]
	\centering
  \includegraphics[width =  \linewidth]{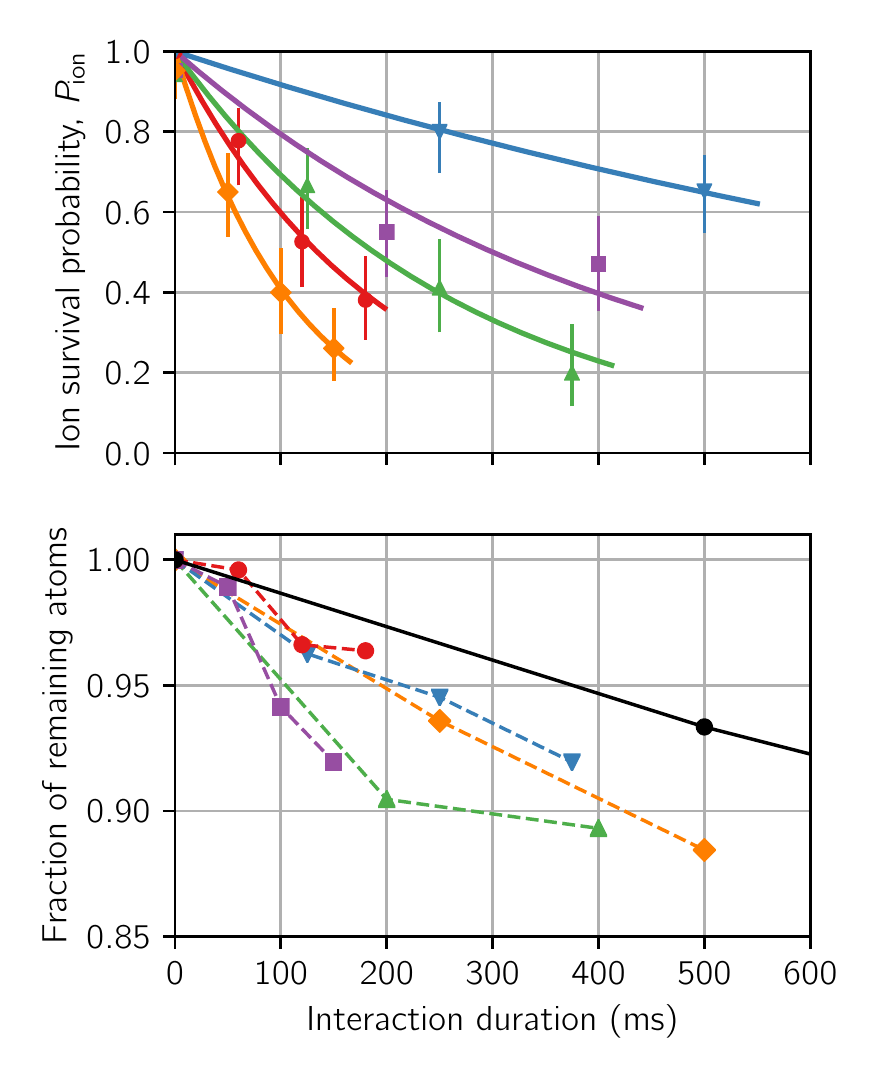}
	\caption{\textbf{Time-dependent Ba$\mathbf{^+}$ and Li loss for variable atomic density $\mathbf{n}$ around \SI{296.31}{\gauss}}.
(upper) Ion survival probability while interacting with the $^6$Li cloud for various atomic densities $n$.
Data points are an average of at least 20 independent experimental realizations.
Error bars denote the upper bound of the $1\sigma$-confidence interval of the underlying binomial distribution.
The solid lines are exponential fits ($e^{-t_{\mathrm{int}}\Gamma_{\mathrm{Loss}}}$) to the respective data.
The fit results and the respective error bars are illustrated as density-dependent loss rate $\Gamma_\mathrm{Loss}(n)$ in fig.~\ref{fig_3}.
(lower) Normalized number of remaining $^6$Li atoms interacting with a single $^{138}$Ba$^+$ ion in dependence on $t_{\mathrm{int}}$.
The markers and the respective colors indicate the association to the data in the upper graph.
Li atoms are removed from the xODT by either spin-changing collisions or elastic atom-ion interactions.
For the presented analysis, we exclude experiments resulting in ion loss, to avoid systematic errors by inelastic collisions.
To mitigate the density uncertainty due to the decay of the atom number, we choose interaction durations resulting in maximal atom loss of $\lesssim \SI{10}{\percent}$.
We further indicate the atom number evolution in absence of interaction (black circles and solid line).
}
	\label{suppl_fig_2}
\end{figure}

\begin{figure}[]
	\centering
  \includegraphics[width =  \linewidth]{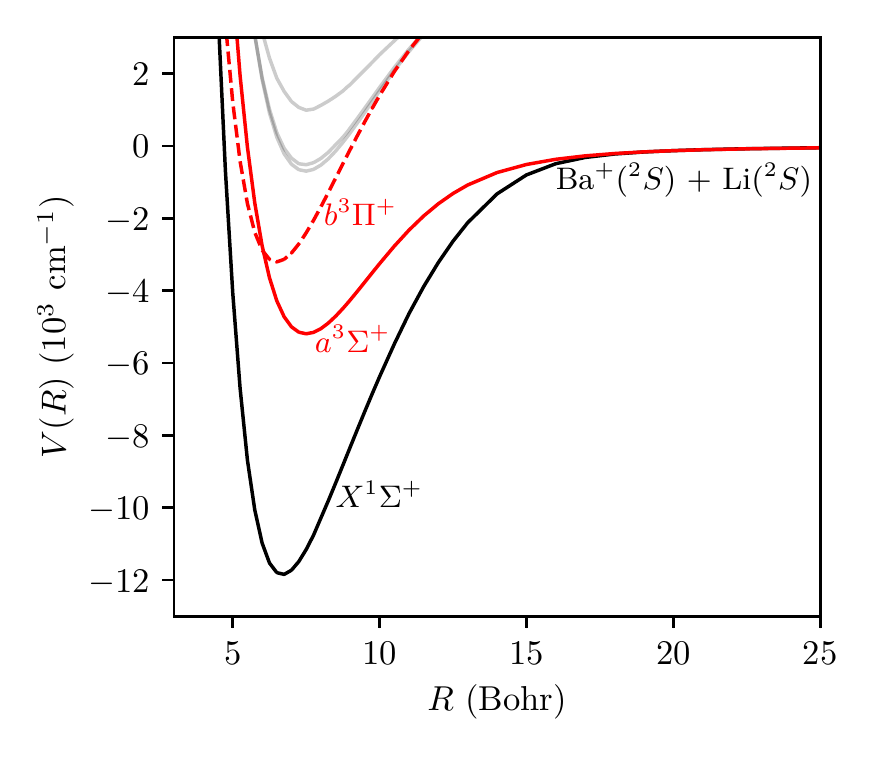}
	\caption{\textbf{Potential energy curves for a Ba$^+$ ion interacting with a Li
atom.} The interaction between ground-state Ba$^+$ ion and Li atom
results in two molecular electronic states of the singlet
$X^1\Sigma^+$ (solid black line) and triplet $a^3\Sigma^+$ (solid red
line) symmetries. The excited molecular electronic state of the
triplet $b^3\Pi$ symmetry (dashed red line) originates from the
interaction of Ba$^+$ ion in the lowest excited $^2D$ state and
ground-state Li atom and crosses the $a^3\Sigma^+$ state at a small
interatomic distance. This crossing combined with spin-orbit coupling
between $a^3\Sigma^+$ and $b^3\Pi$ states results in large
second-order spin-orbit coupling in the collision channels responsible
for the observed Feshbach resonances. Higher-lying excited electronic
states (gray lines) are not directly relevant for the present study.
}
	\label{suppl_fig_3}
\end{figure}

\end{document}